%% file: main.tex
\documentclass{article}
\usepackage{spconf,amsmath,graphicx}
\usepackage{amssymb,amsfonts}
\usepackage{acronym}
\usepackage[caption=false]{subfig}
\usepackage{xcolor}

\acrodef{fl}[FL]{Federated Learning}
\acrodef{sgld}[SGLD]{Stochastic Gradient Langevin Dynamics}
\acrodef{dsgld}[DSGLD]{Decentralized Stochastic Gradient Langevin Dynamics}
\acrodef{awgn}[AWGN]{Additive White Gaussian Noise}
\acrodef{d-wflmc}[D-WFLMC]{Decentralized Wireless Federated Langevin Monte Carlo}
\acrodef{ml}[ML]{Machine Learning}
\acrodef{mcmc}[MCMC]{Markov Chain Monte Carlo}
\acrodef{ps}[PS]{Parameter Server}
\acrodef{noma}[NOMA]{Non Orthogonal Multiple Access}
\acrodef{dnn}[DNN]{Deep Neural Network}
\acrodef{mse}[MSE]{Mean Squared Error}
\acrodef{dsgd}[DSGD]{Decentralized Stochastic Gradient Descent}
\acrodef{ece}[ECE]{Expected Calibration Error}

\title{Channel-driven Decentralized Bayesian Federated Learning for trustworthy decision making in D2D Networks}
\name{Luca Barbieri{$^\star$}, Osvaldo Simeone{$^\dagger$}, and Monica Nicoli{$^\star$} \thanks{The work of O. Simeone was
supported by the European Research Council (ERC) under the European
Union’s Horizon 2020 Research and Innovation Programme (grant agreement
No. 725732) and by an Open Fellowship of the EPSRC.}}
\address{$^\star$ Politecnico di Milano, Milan, Italy\\$^\dagger$ KCLIP lab, Department of Engineering, King's College London}
\begin{document}
%\ninept
%
\maketitle
\begin{abstract}
Bayesian Federated Learning (FL) offers a principled framework to account for the uncertainty caused by limitations in the data available at the nodes \textcolor{black}{implementing collaborative training}. In Bayesian FL, nodes exchange information about local posterior distributions over the model parameters space. 
This paper focuses on Bayesian FL implemented in a device-to-device (D2D) network via Decentralized Stochastic Gradient Langevin Dynamics (DSGLD), a recently introduced gradient-based Markov Chain Monte Carlo (MCMC) method. Based on the observation that DSGLD applies random Gaussian perturbations of model parameters, we propose to leverage channel noise on the D2D links as a mechanism for MCMC sampling. The proposed approach is compared against a conventional implementation of frequentist FL based on compression and digital transmission, highlighting advantages and limitations.  

\end{abstract}
\begin{keywords}
Federated Learning, Markov Chain Monte Carlo, Bayesian inference, Decentralized networks
\end{keywords}

\input{sections/introduction}

\input{sections/system_model}

\input{sections/d-sgld_new}

\input{sections/numerical_results}
\input{sections/conclusions}

\newpage
\bibliographystyle{IEEEtran}
\bibliography{main}

\end{document}

%% file: sections/introduction.tex
\section{Introduction}
\label{sec:introduction}

\ac{fl} enables the collaborative training of \ac{ml} models without the direct exchange of data in both star and fully decentralized architectures  \cite{fl_1,fl_3,fl_4}. 
FL is particularly useful when the participating nodes have limited data. 
This is the case, for instance, in vehicular applications in which individual vehicles can only sense part of a scene (see Fig. 1) \cite{fl_elsevier}. When data sets are size limited, the classical, frequentist, implementation of FL is known to produce models that fail to properly account for the uncertainty of their decisions \cite{simeone2022machine}.  This is an important issue  for safety-critical applications, \textcolor{black}{such as in automated driving services that require trustworthy decisions even in situations with limited data.} 
%requiring trustworthy decisions, including autonomous driving. 
A well-established solution to this problem is to implement Bayesian learning, which encodes uncertainty in the \textcolor{black}{posterior} distribution of the model parameters (see, e.g., \cite{simeone2022machine}). However, a federated implementation of Bayesian learning poses challenges related to the overhead of communicating information about model distributions \cite{d-svgd,jinu2022}.

\begin{figure}
    \centering
    \includegraphics[width=\linewidth]{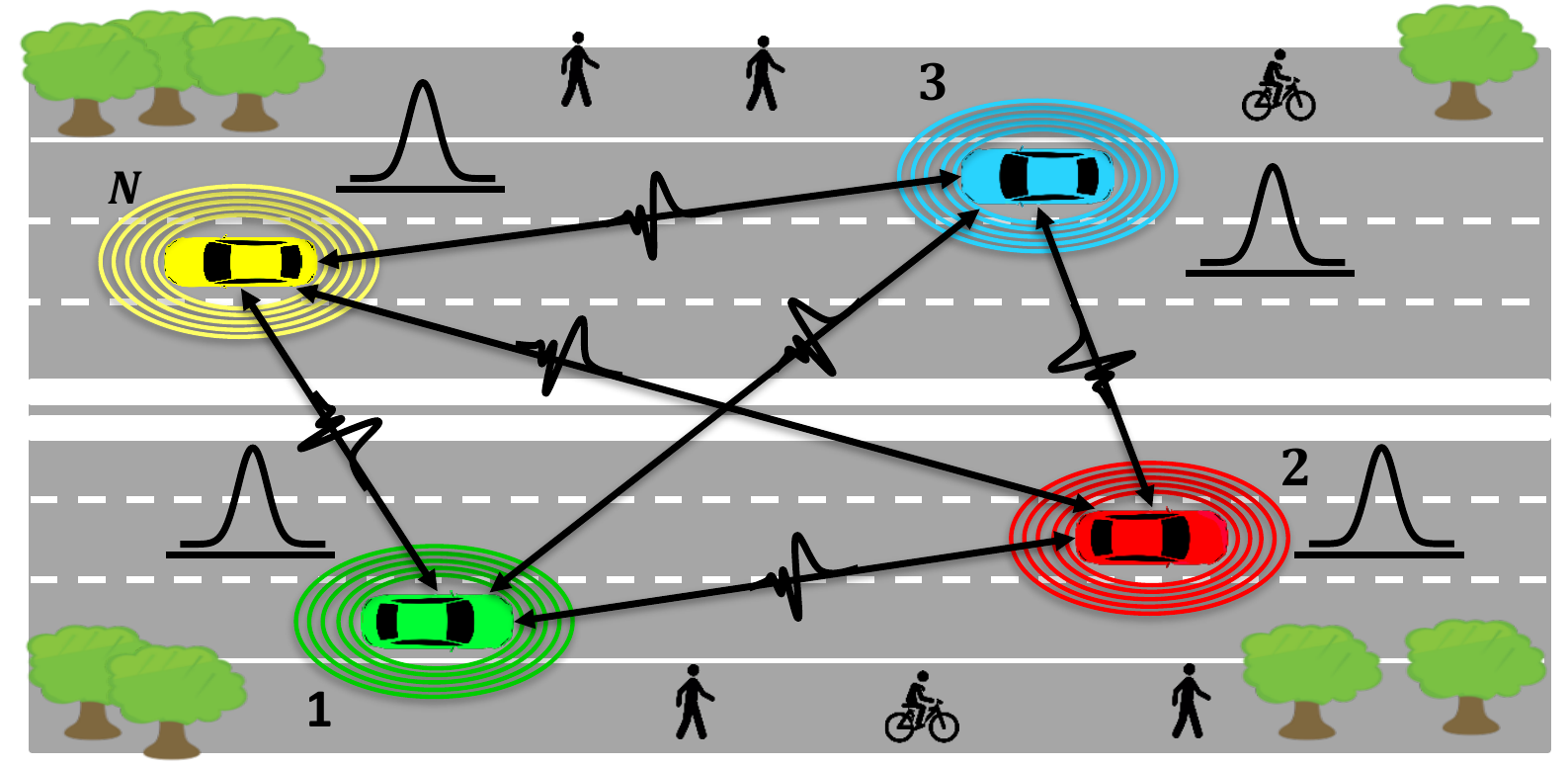}
    %\vspace{-8mm}
    \caption{In the decentralized \ac{fl} set-up under study, nodes (here $N=4$) implement Bayesian learning via Decentralized Stochastic Gradient Langevin Dynamics (DSGLD) by levearging noise on inter-agent communications for sampling.  }
    \label{fig:set-up}
\end{figure}

In a centralized setting, Bayesian learning is practically implemented via approximate methods relying on variational inference (VI) \cite{vi} or  \ac{mcmc}, with the latter representing the target posterior distribution via random samples \cite{mcmc,simeone2022machine}.
Distributed implementations of Bayesian learning have been emerging for both star and device-to-device (D2D) topologies adopting VI \cite{pvi,d-svgd} or \ac{mcmc} \cite{d-sgmcmc,d-sgld}, while assuming ideal communication links. In this paper, we propose a new \ac{mcmc}-based Bayesian FL system tailored for wireless D2D networks with noisy links subject to mutual interference.
%investigate \ac{mcmc}-based Bayesian FL in wireless D2D systems.

To this end, we focus on \ac{sgld} \cite{sgld}, an \ac{mcmc} scheme that has the practical advantage of requiring minor modifications as compared to standard frequentist methods. In fact, SGLD is based on the application of Gaussian perturbations to model parameters updated via gradient descent. Reference \cite{w-flmc} introduced a federated implementation of SGLD over a wireless \emph{star}, i.e., base station-centric, topology. The work \cite{w-flmc} argued that  channel noise between devices and base station can be repurposed to serve as sampling noise for the \ac{sgld} updates, an approach referred to as \emph{channel-driven sampling}. 
%The authors in \cite{w-flmc} exploited also a \ac{noma} scheme to aggregate the samples at a centralized unit, namely a \ac{ps}, over the-air.
In this paper, we draw inspiration from \cite{w-flmc} and study implementations of SGLD in  a \emph{D2D} architecture, as depicted in Fig. \ref{fig:set-up}. We specifically consider the \ac{dsgld} algorithm introduced in \cite{d-sgld} under the assumption of noiseless communications, and we propose an analog communication-based implementation that leverages channel-driven sampling and over-the-air computing \cite{ota_power,ota_1,ota_learning_rate}.
\textcolor{black}{Non-orthogonal multiple access to a shared channel is exploited for over-the-air model aggregation, enabling an efficient \ac{fl} implementation.}
Experiments focus on a challenging automotive use case where vehicles collaboratively train a \ac{dnn} for lidar sensing. 
Numerical results show that the proposed method is able to provide well-calibrated \ac{dnn} models that support trustworthy predictions even when a conventional frequentist \ac{fl} approach based on digital communication fails to meet calibration requirements.

%Over-the-air computing represents an enabling technology in wireless implementations of \ac{fl} procedures as it allows to directly aggregates models exploiting the superposition property of wireless channels via \ac{noma}. 
% Optimization frameworks have been studied under this paradigm to optimize power allocation \
% Nevertheless, channel noise is typically harmful for conventional frequentist \ac{fl} methods.
% Whereas, \ac{mcmc}-based Bayesian \ac{fl} tools can re-use it for sampling. 

The remainder of this paper is organized as follows. 
Sec. \ref{sec:system_model} describes the system model, and Sec. 3 reviews \ac{dsgld}. Sec. \ref{sec:d-sgld-channel} presents the proposed channel-driven method for Bayesian FL, while Sec. \ref{sec:num_results} details the numerical results. 
Finally, Sec. \ref{sec:conclusions} draws some conclusions.

%% file: sections/system_model.tex
%\vspace{-3mm}
\section{System model}
\label{sec:system_model}

We consider the decentralized \ac{fl} system in Fig. \ref{fig:set-up}, which consists of $N$ agents connected according to the undirected graph $\mathcal{G} = (\mathcal{N}, \mathcal{E})$, where $\mathcal{N} = \{1, \ldots, N\}$ is the set of all devices and $\mathcal{E}$ is the set of the  directed edges.
We denote as $\mathcal{N}_k$ the set of neighbors of node $k$ including $k$, while $\mathcal{N}_{\overline{k}}$ is
\textcolor{black}{the same subset excluding node $k$.}
%containing only the neighbors of $k$ (excluding node $k$).
%Each agent is equipped with a single antenna and can communicate with its immediate neighbors through a shared wireless channel
Each agent has access to a local dataset $\mathcal{D}_k = \{(\mathbf{d}_{h}, \ell_h)\}_{h = 1}^{E_k}$ comprising $E_k$ training examples, where $\mathbf{d}_h$ and $\ell_h$ are the input data and the corresponding desired output, or label, respectively. 
Considering all agents, we have the global dataset  $\mathcal{D} = \{\mathcal{D}_k\}_{k = 1}^{N}$.
The goal of the system is to implement Bayesian learning via gradient-based MCMC to obtain a set of samples $\boldsymbol{\theta} \in \mathbb{R}^{m}$ approximating the true global posterior $p(\boldsymbol\theta | \mathcal{D})$ \textcolor{black}{of the model $\boldsymbol{\theta}$ learned cooperatively by all agents.}
%at all agents.
This objective should be met by relying only on local computations at the agents and on D2D communications among the agents.
%on the graph $\mathcal{G}$. 
%direct interactions among the agents. 
\textcolor{black}{The posterior distribution $p(\boldsymbol{\theta}|\mathcal{D})$ describes the uncertainty of the learned model, identifying a set of potential models and the related probabilities.}
%\subsection{Communication Model}

%\textcolor{red}{Add here communication model }

%The agents are assumed to communicate over full-duplex flat-fading channels where the channels coefficients remain constant throughout the overall communication block and may vary across successive blocks, possibly in a correlated way.
The agents communicate over full-duplex shared \textcolor{black}{wireless} channels impaired by \ac{awgn}.
Communication is organized into blocks, each consisting of $m$ channel uses. 
%We assume block flat-fading channels, where the channels coefficients remain constant throughout the overall communication block and may vary over successive blocks, possibly in a correlated way. 
%Each block contains $m$ channel uses, allowing the transmission of the local samples by a device. 
In each block $s = 1, 2, \ldots$  the $m$-sample signal received by  agent $k$ can be expressed as
\begin{equation}
    \mathbf{y}_k^{[s]} = \sum_{j \in \mathcal{N}_{\overline{k}}} \mathbf{x}_j^{[s]} +  \mathbf{z}_k^{[s]}\, ,  \label{eq:received_signal}
\end{equation} where $\mathbf{z}_k^{[s]} \sim \mathcal{N}(0,N_0 \mathbf{I}_m)$ is the channel noise ($\mathbf{I}_m$ is the $m \times m$ identity matrix), and the $m$-sample block $\mathbf{x}_j^{[s]}$ transmitted  by each node $j$ satisfies the power constraint $\| \mathbf{x}_{j}^{[s]}\|^{2} \leq m P$.

\section{Distributed Stochastic Gradient Langevin Dynamics}
In this section, we review \ac{dsgld} \cite{d-sgld}, which applies to a system with ideal inter-agent communication.

\subsection{Centralized Stochastic Gradient Langevin Dynamics}

We start by reviewing the standard SGLD scheme \cite{sgld}, which applies in the ideal case where the global data set $\mathcal{D}$ is available at a central processor. Given a likelihood function \textcolor{black}{$p(\mathcal{D}_k| \boldsymbol\theta)$}  describing the shared ML model adopted by the agents (e.g., a neural network), and a prior distribution $p(\boldsymbol\theta)$, the \emph{global posterior distribution} is defined as 
\begin{equation}\label{eq:post}
    p(\boldsymbol\theta | \mathcal{D}) \propto p(\boldsymbol\theta) \prod_{k = 1}^{N} p(\mathcal{D}_k | \boldsymbol\theta) \, , 
\end{equation}
with $p(\mathcal{D}_k | \boldsymbol\theta) = \prod_{h = 1}^{E_k} p(\ell_h | \mathbf{d}_h, \boldsymbol\theta)$. \ac{sgld} produces samples whose asymptotic distribution approximately matches the global posterior $p(\boldsymbol\theta | \mathcal{D})$.

This is accomplished by adding Gaussian noise to standard gradient descent updates via the following iterative update rule \cite{sgld}
\begin{equation}
    \boldsymbol\theta^{[s+1]} = \boldsymbol\theta^{[s]} - \eta \nabla f(\boldsymbol\theta^{[s]}) + \sqrt{2\eta} \boldsymbol\xi^{[s+1]} \, , 
    \label{eq:sgld}
\end{equation}
where $s = 1, 2, \ldots; $ $\eta$ is the step size; $f (\boldsymbol\theta) = \sum_{k = 1}^{N} f_k(\boldsymbol\theta)$ is the negative logarithm of the unnormalized global posterior $p(\boldsymbol\theta) \prod_{k = 1}^{N} p(\mathcal{D}_k | \boldsymbol\theta)$,  with 
\begin{equation}
    f_k(\boldsymbol\theta) = -\log p(\mathcal{D}_k | \boldsymbol{\theta}) - \dfrac{1}{N} \log p(\boldsymbol\theta) \, ,  \label{eq:gradient}
\end{equation} 
and $\boldsymbol\xi^{[s+1]}$ is a sequence of identical and independent (i.i.d.) random vectors following the Gaussian distribution $\mathcal{N}(0, \mathbf{I}_{m})$, independent of the initialization $\boldsymbol\theta^{[0]} \in \mathbb{R}^{m}$.

% Equivalently, the global posterior can be computed considering the local subposteriors $\widetilde{p}(\boldsymbol\theta | \mathcal{D}_k) \propto p(\boldsymbol\theta)^{1/K} p(\mathcal{D}_k | \boldsymbol\theta)$ at the agents as
% \begin{equation}
%     p(\boldsymbol\theta | \mathcal{D}) \propto \prod_{k=1}^{N} \widetilde{p}(\boldsymbol\theta | \mathcal{D}_k) \, , 
% \end{equation}
% with $ \widetilde{p}(\boldsymbol\theta | \mathcal{D}_k) \propto p(\boldsymbol\theta)^{1/K} p(\mathcal{D}_k | \boldsymbol\theta)$.

%In contrast, \ac{sghmc} utilizes the underdamped Langevin diffusion process, producing samples according to the following update rules 
%\begin{align}
%    \boldsymbol\nu^{[s+1]} &= \boldsymbol\nu^{[s]} - \eta [\gamma \boldsymbol\nu^{[s]} + \nabla f(\boldsymbol\theta^{[s]})] + \sqrt{2 \gamma \eta} \boldsymbol\xi^{[s+1]} \, ,  \\ 
%    \boldsymbol\theta^{[s+1]} &= \boldsymbol\theta^{[s]} + \eta \boldsymbol\nu^{[s+1]} \, ,
%    \label{eq:sghmc}
%\end{align}
%where $\gamma$ is the momentum rate; $\nabla f(\boldsymbol\theta^{[s]})$ is defined as in \eqref{eq:gradient}; and $\boldsymbol\xi^{[s+1]} \sim \mathcal{N}(0,\mathbf{I}_m)$ is a sequence of zero-mean i.i.d. Gaussian distributed random vectors independent of the initialization $\boldsymbol\theta^{[0]} \in \mathbb{R}^{m}$ and $\boldsymbol\nu^{[0]} \in \mathbb{R}^{m}$.

% 

\subsection{Decentralized Stochastic Gradient Langevin Dynamics}

\ac{dsgld} \cite{d-sgld} is an extension of SGLD that applies to D2D networks. In DSGLD,  each agent $k$ applies the following update rule 
\begin{equation}
    \boldsymbol\theta_k^{[s+1]} = \sum_{j \in \mathcal{N}_k} w_{kj} \boldsymbol\theta_j^{[s]} - \eta \nabla f_k(\boldsymbol\theta_{k}^{[s]}) + \sqrt{2 \eta} \boldsymbol\xi_k^{[s+1]} \, , \label{eq:dsgld}
\end{equation} 
where $w_{kj}$ is the $(k,j)$-th entry of a symmetric, doubly stochastic matrix $\mathbf{W}$. 
% \begin{equation}
%     \nabla f_k (\boldsymbol\theta_{k}^{[s]}) = - \nabla \log p(\mathcal{D}_{k} | \boldsymbol\theta_k^{[s]}) - \dfrac{1}{N} \nabla \log p( \boldsymbol\theta_k^{[s]}) \, ,
% \end{equation}
% is the local gradient. 
%Similarly, \ac{d-sghmc} \cite{JMLR:v22:21-0307} lets each agent $k$ update its local samples as
%\begin{align}
%    \boldsymbol\nu_k^{[s+1]} &= \boldsymbol\nu_k^{[s]} - \eta [\gamma \boldsymbol\nu_k^{[s]} + \nabla f_k(\boldsymbol\theta_k^{[s]})] + \sqrt{2 \gamma \eta} \boldsymbol\xi_k^{[s+1]} \, ,  \\ 
%    \boldsymbol\theta_k^{[s+1]} &= \sum_{j \in \mathcal{N}_k} w_{kj} \boldsymbol\theta_j^{[s]} + \eta \boldsymbol\nu_k^{[s+1]} \, ,
%    \label{eq:dsgmhc}
%\end{align}
%\begin{equation}
%    \boldsymbol\theta_{k}^{[s+1]} = \sum_{j \in \mathcal{N}_{k}}w_{kj}\boldsymbol\theta_{j}^{[s]} + (1 - \gamma)\eta \boldsymbol\nu_{k}^{[s]}  - \eta \nabla f_k(\boldsymbol\theta^{[s]}) + \sqrt{2 \gamma \eta} \boldsymbol\xi_{k}^{[s+1]} \, , \label{eq:dsgmhc}
%\end{equation}
Accordingly, in DSGLD, at each iteration $s$, each agent $k$ combines the current model iterates $\boldsymbol\theta_j^{[s]}$ from its neighbors $j\in \mathcal{N}_k$, and it also applies a noisy gradient update as in the SGLD update (\ref{eq:sgld}). Under a properly chosen learning rate $\eta$ and assuming  graph $\mathcal{G}$ to be connected and the functions $f_k(\cdot)$ to be smooth and strongly convex, the distributions of the samples produced by DSGLD converge to the global posterior (\ref{eq:post}) \cite{d-sgld}. In the following, we will set the weights $w_{kj}$ to be equal for all nodes, i.e., $w_{kj}=w\in(0,1/\max_{k \in \mathcal{N}} |\mathcal{N}_k| ]$ and $w_{kk}=1-|\mathcal{N}_k|w$, in order to simplify the wireless implementation.
% \begin{equation}
%     w_{k,j} = 
%     \begin{cases}
%     \gamma & \forall k \in \mathcal{N}, \forall j \in \mathcal{N}_k \\
%     1 - |\mathcal{N}_k|\gamma  & \forall \left\{ k,j \right\} \in \mathcal{N} \, \text{with} \, k = j \\
%     0 &  \text{otherwise}
%     \end{cases}
% \end{equation}
% Parameter $\gamma$ generally belongs to the set $[0, 1/\mathcal{N}_k]$ and it is here chosen as $\gamma = 2/\lambda_1(\mathbf{L} + \lambda_{N-1}(\mathbf{L}))$ where $\mathbf{L}$ is the Laplacian matrix while $\lambda_{\ell}(\mathbf{L})$ denotes the $\ell$-th eigenvalue of $\mathbf{L}$.

% The learning protocol for \ac{d-sgld} runs for a number of communication rounds $s = 1, 2, \ldots $ and alternates local computations at the agents to estimate the gradient $\nabla f_k (\boldsymbol\theta_{k}^{[s]})$ and the aggregation of the samples $ \{\boldsymbol\theta_j^{[s]} \}_{j \in \mathcal{N}_k}$ received by neighbors to obtain the newly updated samples of the (approximated) global posterior. 

%% file: sections/d-sgld_new.tex
\section{Channel-Driven Decentralized Bayesian Federated Learning}
\label{sec:d-sgld-channel}
In this section, we propose an implementation of DSGLD that leverages channel-driven sampling and over-the-air computing via analog transmission, \textcolor{black}{referred} to as CD-DSGLD.  

\subsection{CD-DSGLD}

Using full-duplex radios, in CD-DSGLD, all agents transmit simultaneously in each block $s$ of $m$ channel uses. Block $s$ is used to exchange information required for the application of the $s$-th D-SGLD update (\ref{eq:dsgld}). Accordingly, the transmitted signal $\mathbf{x}_{j}^{[s]}$ is an uncoded function of the local iterate $\boldsymbol\theta_j^{[s]}$
\begin{equation}
    \mathbf{x}_j^{[s]} = w \alpha_j^{[s]} \boldsymbol\theta_{j}^{[s]} \, , 
    \label{eq:encoded_functions}
\end{equation}
with $\alpha_{j}^{[s]}$ being a power control parameter.  Given the received signal \eqref{eq:received_signal}, each device $k \in \mathcal{N}$ applies the update
\begin{equation}
    \hat{\boldsymbol\theta}_k^{[s+1]} = w_{kk} \boldsymbol\theta_k^{[s]} + \dfrac{\mathbf{y}_{k}^{[s]}}{\beta^{[s]}} - \eta \nabla f_k(\boldsymbol{\theta}_k^{[s]}),
    \label{eq:channel_driven_dsgld}
\end{equation}
where $\beta^{[s]}$ is a receiver-side scaling factor.
%As we discuss next, the iteration (\ref{eq:channel_driven_dsgld}) is an estimate of the DSGLD update (\ref{eq:dsgld}).

By plugging \eqref{eq:received_signal} and \eqref{eq:encoded_functions} into \eqref{eq:channel_driven_dsgld}, we obtain \textcolor{black}{the update rule}
\begin{equation}
    \hat{\boldsymbol\theta}_k^{[s+1]} = w_{kk} \boldsymbol\theta_k^{[s]} + \sum_{j \in \mathcal{N}_{\overline{k}}} \dfrac{w \alpha_j^{[s]} \boldsymbol\theta_j^{[s]}}{\beta^{[s]}} - \eta \nabla f_k(\boldsymbol{\theta}_k^{[s]}) + \dfrac{\mathbf{z}_k^{[s]}}{\beta^{[s]}} \, , 
    \label{eq:prop_scheme}
\end{equation}
\textcolor{black}{which} equals 
%Therefore, the update in \eqref{eq:prop_scheme} equals
the \ac{dsgld} update (\ref{eq:dsgld}) if (\emph{i}) the noise introduced by the channel has variance $2 \eta$, i.e., if the receiver scaling factor is selected as $\beta^{[s]} = \sqrt{N_0/(2 \eta)}$;
and (\emph{ii}) if the power scaling factor is chosen as  $\alpha_j^{[s]} = \beta^{[s]}$. However, condition (\emph{ii}) cannot be met in general due to the power constraints. 
% However, owning to the different power control parameters $\alpha_j^{[s]}$ originating from different $\| \boldsymbol{\theta}_j^{[s]} \|^{2}$ values, the update in \eqref{eq:channel_driven_dsgld} provides only an approximation to the true \ac{d-sgld} update in \eqref{eq:dsgld}.

% The first term $\Delta \mathbf{g}_k^{[s]}$ denotes an error introduced when the samples are aggregated at the receiver, while the second one $\Delta \mathbf{n}_k^{[s]} $ indicates a mismatch between the noise power required for the sampling process of \ac{d-sgld} and the noise power introduced from the channel. 

\subsection{Optimization of the Scaling Factors}
\label{subsec:optimization}

To minimize the discrepancy between \eqref{eq:dsgld} and \eqref{eq:channel_driven_dsgld}, we  propose to jointly optimize the scaling factors $\{\alpha_j^{[s]}\}_{j \in \mathcal{N}}$, while setting $\beta^{[s]} = \sqrt{N_0/(2 \eta)}$. To this end, we set up the problem
\begin{equation}
\begin{aligned}
\min_{\{\alpha_j^{[s]} > 0\}_{j \in \mathcal{N}}} &  \left(\hat{\boldsymbol{\theta}}_{k}^{[s+1]} - \boldsymbol{\theta}_{k}^{[s+1]}\right)^2,  \\
\textrm{s.t.} \qquad &  \|\mathbf{x}_j^{[s]} \|^2 \leq m P \qquad \forall j \in \mathcal{N},\\
\end{aligned}  
\label{eq:obj_function}
\end{equation}
with 
\begin{equation}
    \hat{\boldsymbol{\theta}}_{k}^{[s+1]} - \boldsymbol{\theta}_{k}^{[s+1]} = \dfrac{w}{\beta^{[s]}}
    \sum_{j \in \mathcal{N}_{\overline{k}}} \alpha_j^{[s]} \boldsymbol\theta_j^{[s]} - w \sum_{j \in \mathcal{N}_{\overline{k}}} \boldsymbol{\theta}_j^{[s]} \, ,
\end{equation}
Therefore, problem \eqref{eq:obj_function} is a quadratic program that  can be solved using standard tools.

%% file: sections/numerical_results.tex
\subsection{Benchmark Quantized Digital Frequentist Implementation}

As a benchmark, we adopt a conventional digital implementation of a frequentist \ac{fl} based on compression and \ac{dsgd} \cite{dsgd}. DSGD implements the update rule \eqref{eq:dsgld} by removing the Gaussian noise, i.e., setting $\boldsymbol\xi_k^{[s+1]} = 0$. To implement DSGD using digital transmission, for each iteration, and corresponding communication block $s$, we assume communication over the channel \eqref{eq:received_signal} via non-orthogonal access with receivers treating interference as noise \cite{tin_noma}. 

Accordingly, each (full-duplex) node $k$ can communicate up to $m \log_2 \left(1 + \text{SINR}_k \right)$ bits per block, where $\text{SINR}_k = P/((|\mathcal{N}_k| -1)P + N_0)$ is the Signal-to-Interference-plus-Noise Ratio (SINR) at node $k$. 
Each transmitter applies stochastic quantization \cite{qsgd} and top-$t$ sparsification as in \cite{top-k}. 
Accordingly, using $N_b = 10$ bits to encode each entry of the \ac{ml} model parameter vector $\boldsymbol{\theta}_k^{[s]}$, the number of bits per block is given by $\log_2 {m \choose t} + t N_b$, where $\log_2 {m \choose t}$ denotes the overhead required for encoding the top-$t$ numbers using Golomb position encoding \cite{golomb_encoding}. Therefore, the parameter $t$ is selected as the smallest integer value $t$ such this number of bits can be communicated in a block, i.e., such that the inequality
\begin{equation}
  m \log_2 \left(1 + \text{SINR}_k \right) \geq \log_2 {m \choose t} + t N_b
\end{equation} holds. We refer to the benchmark scheme as Quantized DSGD (Q-DSGD).

\section{Numerical results}
\label{sec:num_results}

\begin{figure}[!t]
    \centering
    \subfloat[\label{fig:accuracy}]{
    \includegraphics[width=0.99\linewidth]{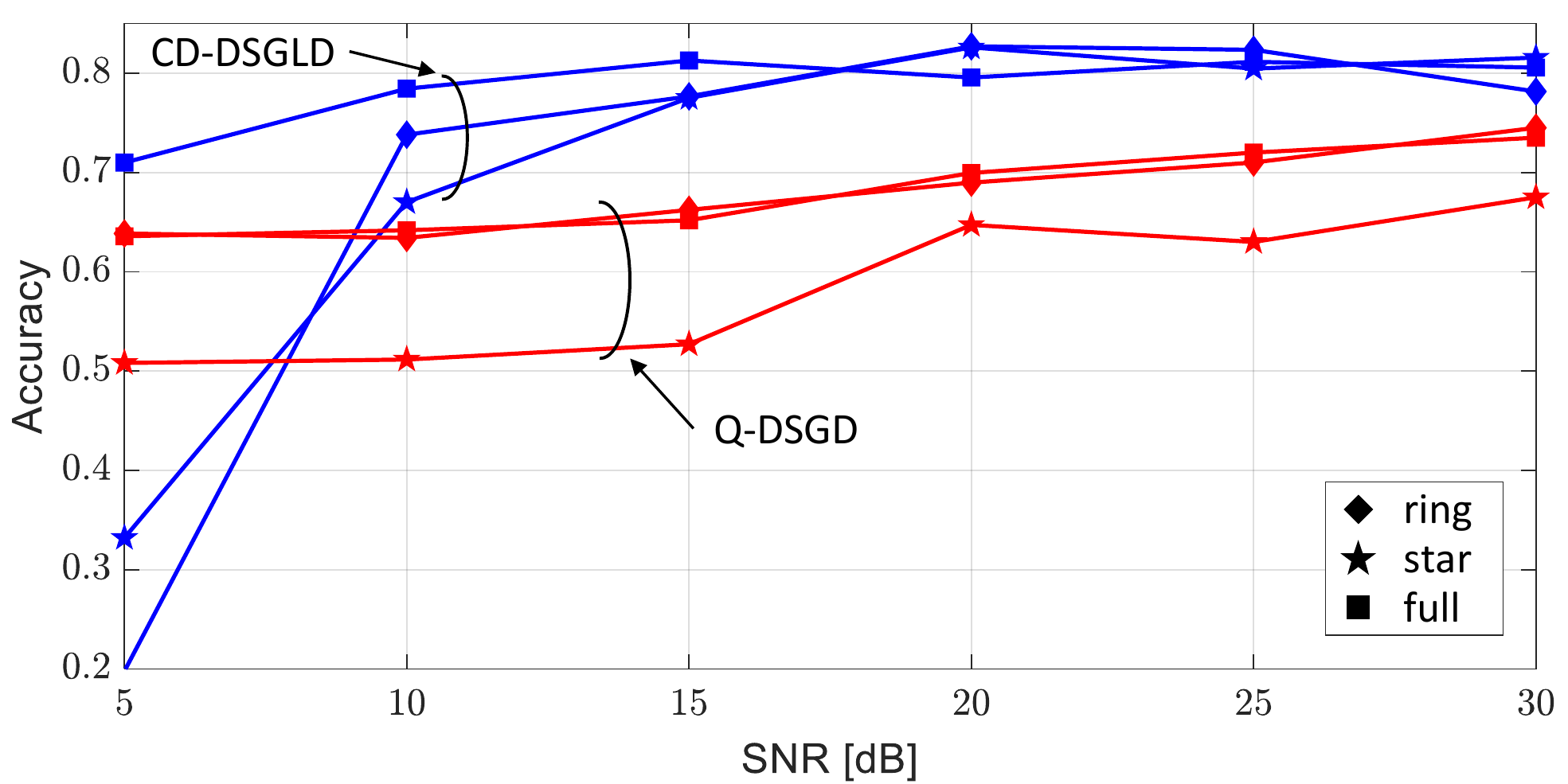}
    }
    
    %\vspace{-3mm}
    \subfloat[\label{fig:ece}]{
    \includegraphics[width=0.99\linewidth]{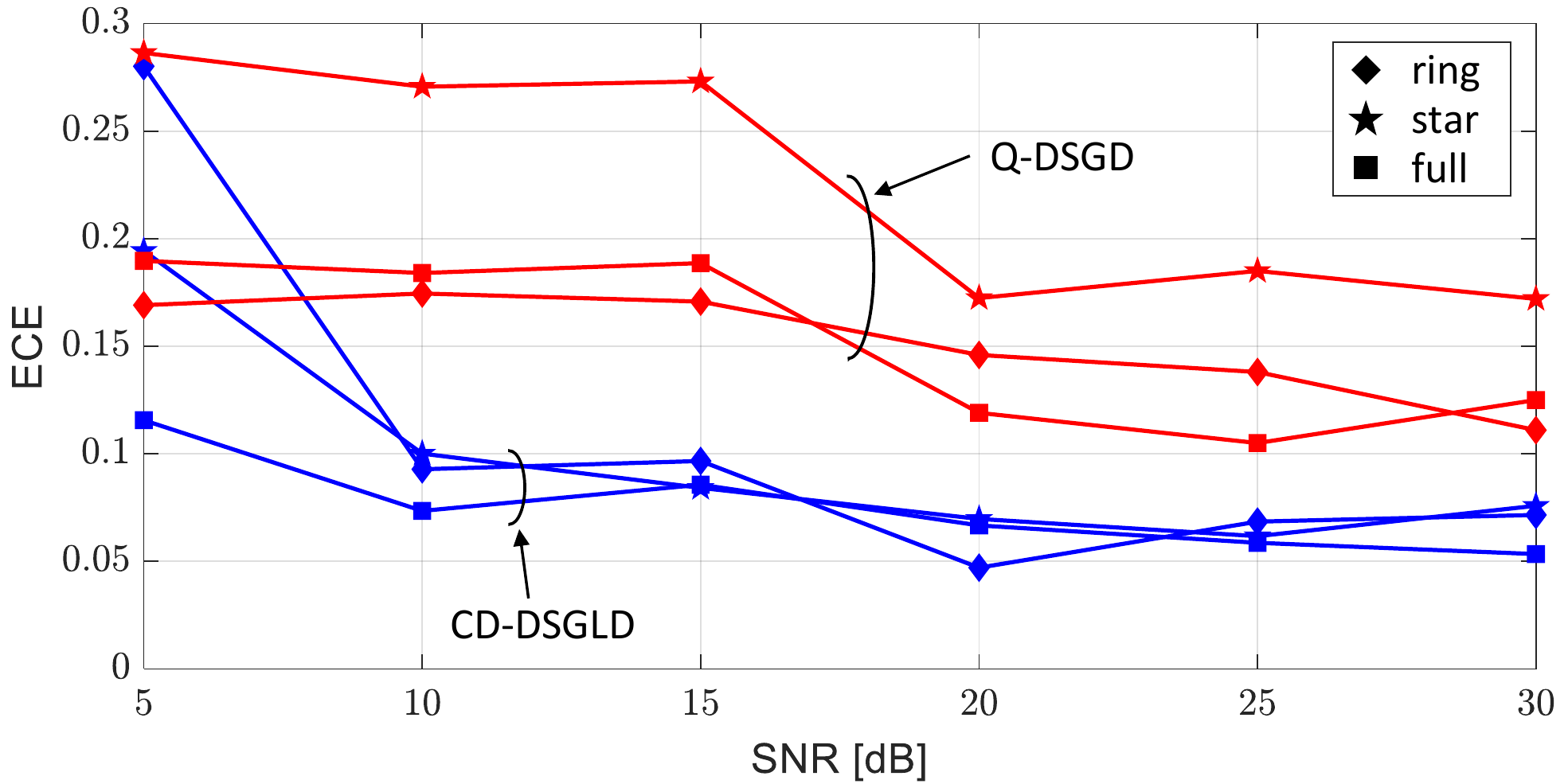}
    }
    %\vspace{-3mm}
    \caption{Accuracy (a) and ECE (b) for the proposed CD-DSGLD scheme and for the conventional Q-DSGD as a function of the SNR for \textcolor{black}{full, ring, and star topologies.}}
    \label{fig:ece_comparison}
\end{figure}

This section presents experiments used to evaluate the performance of CD-SGLD against the conventional digital implementation of DSGD, Q-DSGD, reviewed in the previous section. 

\subsection{Simulation Setting}
%\vspace{-10mm}
We consider a cooperative sensing \textcolor{black}{use case}, where $N = 5$  vehicles collaborate to detect and classify 6 different road users/objects from point cloud data \textcolor{black}{gathered by on-board lidar} as in \cite{fl_elsevier}.
\textcolor{black}{To infer the road object categories, vehicles rely on a PointNet \ac{dnn} implemented as in \cite{fl_elsevier}, with $m = 40 \text{,}855$ parameters and $40$ trainable layers.}
%To solve this task, vehicles rely on a PointNet \ac{dnn} to infer the road object categories.
%The PointNet architecture is implemented as in \cite{fl_elsevier}, containing $m = 40 \text{,}855$ parameters and $40$ trainable layers. 
The agents hold a training dataset composed by $40$ examples for each one of the 6 classes, and the performances are evaluated over a separated validation dataset comprised by $2\text{,}400$ examples evenly partitioned across the 6 classes. 
We consider three different connectivity patterns for assessing the performances: (\emph{i}) a fully connected topology, (\emph{ii}) a star topology and (\emph{iii}) a ring topology.

\subsection{Implementation}

For both CD-DSGLD and Q-DSGD, we set the learning rate to $\eta = 10^{-4}$, and the total number of iterations to $s = 15\text{,}000$. For CD-SGLD, we follow the standard approach of discarding the first samples produced during the ``burn-in'' period \cite{w-flmc}. The burn-in period is is set to $14,900$ iterations. Furthermore, we adopt a standard Gaussian $ \mathcal{N}(0, \mathbf{I}_m)$ as the prior $p(\boldsymbol\theta)$.
For both methods, we set $w = 2/\lambda_1(\mathbf{L} + \lambda_{N-1}(\mathbf{L}))$, where $\mathbf{L}$ is the Laplacian matrix of the graph $\mathcal{G} = (\mathcal{N}, \mathcal{E})$, while $\lambda_{\ell}(\mathbf{L})$ denotes the $\ell$-th eigenvalue of $\mathbf{L}$ \cite{Hong:21}.

\subsection{Performance Metrics}
As performance metrics, we consider the standard measure of test accuracy and the \ac{ece}, with the latter quantifying the ability of the model to provide reliable uncertainty measures \cite{ece}.
The \ac{ece} is evaluated based on the confidence level produced by the model as the maximum probability assigned by the last, softmax, layer to the possible outputs. The ECE measures how well the confidence levels reflect the true accuracy of the decision corresponding to the maximum probability, and it is computed as follows. 
At first, the test set is divided into a set of $T$ bins $\{B_t\}_{t=1}^{T}$ and for each bin $B_t$ the average accuracy $\text{acc}(B_t)$ and confidence $\text{conf}(B_t)$ are computed.
Then, the \ac{ece} is computed by taking into account all accuracy/confidence values for all bins as \cite{ece}
\begin{equation}
\text{ECE} = \sum_{t=1}^{T} \dfrac{|B_t|}{n} |\text{acc}(B_t) - \text{conf}(B_t)| \, , 
\end{equation}
where $|B_t|$ and $n$ denote the number of examples in the $t$-th bin and the overall number of examples in the validation set, respectively. 

\subsection{Results}
Fig. 2 reports the test accuracy and the \ac{ece} of CD-DSGLD and Q-DSGD  as a function of the SNR, defined as $\mathrm{SNR}=P/N_0$. 
The main conclusion from the figure is that Bayesian FL via CD-DSGLD can significantly enhance the calibration as compared to frequentist FL implemented via Q-DSGD, as long as the SNR is sufficiently large. The advantages of CD-DSLGD in terms of calibration come at the cost of a lower accuracy at low values of the SNR if the connectivity of the network is reduced. In this case, the excessive channel noise does not match well the requirement of DSGLD, causing a performance loss for CD-SGLD. However, for sufficiently large SNR, or for a well-connected network at all SNRs, CD-DSGLD outperforms Q-DSGD both in terms of accuracy and calibration. 

%% file: sections/conclusions.tex
\section{Conclusions}
\label{sec:conclusions}

This paper has proposed a novel channel-driven Bayesian \ac{fl} strategy for decentralized, or D2D networks \textcolor{black}{where over-the-air computing is exploited to aggregate the samples during the wireless propagation}. 
The experimental results \textcolor{black}{considered a challenging cooperative sensing task for automotive applications, and }confirmed the superior performance of the proposed method in providing more accurate uncertainty quantification  as compared to a frequentist \ac{fl} method based on standard  digital transmission of quantized model parameters. 
Future works will target the integration of more complex channel models.  

%% file: main.bbl
\begin{thebibliography}{10}
\providecommand{\url}[1]{#1}
\csname url@samestyle\endcsname
\providecommand{\newblock}{\relax}
\providecommand{\bibinfo}[2]{#2}
\providecommand{\BIBentrySTDinterwordspacing}{\spaceskip=0pt\relax}
\providecommand{\BIBentryALTinterwordstretchfactor}{4}
\providecommand{\BIBentryALTinterwordspacing}{\spaceskip=\fontdimen2\font plus
\BIBentryALTinterwordstretchfactor\fontdimen3\font minus
  \fontdimen4\font\relax}
\providecommand{\BIBforeignlanguage}[2]{{%
\expandafter\ifx\csname l@#1\endcsname\relax
\typeout{** WARNING: IEEEtran.bst: No hyphenation pattern has been}%
\typeout{** loaded for the language `#1'. Using the pattern for}%
\typeout{** the default language instead.}%
\else
\language=\csname l@#1\endcsname
\fi
#2}}
\providecommand{\BIBdecl}{\relax}
\BIBdecl

\bibitem{fl_1}
H.~B. McMahan, E.~Moore, D.~Ramage \emph{et~al.}, ``Communication-efficient
  learning of deep networks from decentralized data,'' \emph{arXiv e-prints},
  2016.

\bibitem{fl_3}
T.~Li, A.~K. Sahu, A.~Talwalkar, and V.~Smith, ``Federated learning:
  Challenges, methods, and future directions,'' \emph{IEEE Signal Processing
  Magazine}, vol.~37, no.~3, pp. 50--60, 2020.

\bibitem{fl_4}
S.~Savazzi, M.~Nicoli, M.~Bennis \emph{et~al.}, ``Opportunities of federated
  learning in connected, cooperative, and automated industrial systems,''
  \emph{IEEE Communications Magazine}, vol.~59, no.~2, pp. 16--21, 2021.

\bibitem{fl_elsevier}
L.~Barbieri, S.~Savazzi, M.~Brambilla, and M.~Nicoli, ``Decentralized federated
  learning for extended sensing in 6g connected vehicles,'' \emph{Vehicular
  Communications}, p. 100396, 2021.

\bibitem{simeone2022machine}
O.~Simeone, \emph{{Machine Learning for Engineers}}.\hskip 1em plus 0.5em minus
  0.4em\relax Cambridge University Press, 2022.

\bibitem{d-svgd}
R.~Kassab and O.~Simeone, ``Federated generalized bayesian learning via
  distributed stein variational gradient descent,'' \emph{IEEE Transactions on
  Signal Processing}, vol.~70, pp. 2180--2192, 2022.

\bibitem{jinu2022}
J.~Gong, O.~Simeone, and J.~Kang, ``Compressed particle-based federated
  bayesian learning and unlearning,'' \emph{arXiv e-prints}, 2022.

\bibitem{vi}
D.~M. Blei, A.~Kucukelbir, and J.~D. McAuliffe, ``Variational inference: A
  review for statisticians,'' \emph{Journal of the American Statistical
  Association}, vol. 112, no. 518, pp. 859--877, 2017.

\bibitem{mcmc}
E.~{Angelino}, M.~J. {Johnson}, and R.~P. {Adams}, ``{Patterns of Scalable
  Bayesian Inference},'' \emph{arXiv e-prints}, Feb. 2016.

\bibitem{pvi}
M.~Ashman, T.~D. Bui, C.~V. Nguyen \emph{et~al.}, ``Partitioned variational
  inference: A framework for probabilistic federated learning,'' \emph{CoRR},
  vol. abs/2202.12275, 2022.

\bibitem{d-sgmcmc}
S.~Ahn, B.~Shahbaba, and M.~Welling, ``Distributed stochastic gradient mcmc,''
  in \emph{Proceedings of the 31st International Conference on Machine
  Learning}, ser. Proceedings of Machine Learning Research, E.~P. Xing and
  T.~Jebara, Eds.\hskip 1em plus 0.5em minus 0.4em\relax Bejing, China: PMLR,
  22--24 Jun 2014, pp. 1044--1052.

\bibitem{d-sgld}
M.~Garbazbalaban, X.~Gao, Y.~Hu, and L.~Zhu, ``{Decentralized Stochastic
  Gradient Langevin Dynamics and Hamiltonian Monte Carlo},'' \emph{Journal of
  Machine Learning Research}, vol.~22, no. 239, pp. 1--69, 2021.

\bibitem{sgld}
M.~Welling and Y.~W. Teh, ``Bayesian learning via stochastic gradient langevin
  dynamics,'' in \emph{Proceedings of the 28th International Conference on
  Machine Learning}, ser. ICML'11.\hskip 1em plus 0.5em minus 0.4em\relax
  Omnipress, 2011, p. 681–688.

\bibitem{w-flmc}
D.~{Liu} and O.~{Simeone}, ``{Wireless Federated Langevin Monte Carlo:
  Repurposing Channel Noise for Bayesian Sampling and Privacy},'' \emph{arXiv
  e-prints}, Aug. 2021.

\bibitem{ota_power}
X.~Cao, G.~Zhu, J.~Xu \emph{et~al.}, ``Optimized power control design for
  over-the-air federated edge learning,'' \emph{IEEE Journal on Selected Areas
  in Communications}, vol.~40, no.~1, pp. 342--358, 2022.

\bibitem{ota_1}
K.~Yang, T.~Jiang, Y.~Shi, and Z.~Ding, ``Federated learning via over-the-air
  computation,'' \emph{IEEE Transactions on Wireless Communications}, vol.~19,
  no.~3, pp. 2022--2035, 2020.

\bibitem{ota_learning_rate}
C.~Xu, S.~Liu, Z.~Yang \emph{et~al.}, ``Learning rate optimization for
  federated learning exploiting over-the-air computation,'' \emph{IEEE Journal
  on Selected Areas in Communications}, vol.~39, no.~12, pp. 3742--3756, 2021.

\bibitem{dsgd}
R.~{Xin}, S.~{Kar}, and U.~A. {Khan}, ``{An introduction to decentralized
  stochastic optimization with gradient tracking},'' \emph{arXiv e-prints},
  Jul. 2019.

\bibitem{tin_noma}
W.~Shin, M.~Vaezi, B.~Lee \emph{et~al.}, ``Non-orthogonal multiple access in
  multi-cell networks: Theory, performance, and practical challenges,''
  \emph{IEEE Communications Magazine}, vol.~55, no.~10, pp. 176--183, 2017.

\bibitem{qsgd}
D.~{Alistarh}, D.~{Grubic}, J.~{Li} \emph{et~al.}, ``{QSGD:
  Communication-Efficient SGD via Gradient Quantization and Encoding},''
  \emph{arXiv e-prints}, Oct. 2016.

\bibitem{top-k}
S.~{Shi}, X.~{Chu}, K.~C. {Cheung}, and S.~{See}, ``{Understanding Top-k
  Sparsification in Distributed Deep Learning},'' \emph{arXiv e-prints}, Nov.
  2019.

\bibitem{golomb_encoding}
F.~Sattler, S.~Wiedemann, K.~Müller, and W.~Samek, ``Sparse binary
  compression: Towards distributed deep learning with minimal communication,''
  in \emph{2019 International Joint Conference on Neural Networks (IJCNN)},
  2019, pp. 1--8.

\bibitem{Hong:21}
H.~Xing, O.~Simeone, and S.~Bi, ``Federated learning over wireless
  device-to-device networks: Algorithms and convergence analysis,'' \emph{IEEE
  Journal on Selected Areas in Communications}, vol.~39, no.~12, pp.
  3723--3741, 2021.

\bibitem{ece}
C.~{Guo}, G.~{Pleiss}, Y.~{Sun}, and K.~Q. {Weinberger}, ``{On Calibration of
  Modern Neural Networks},'' \emph{arXiv e-prints}, Jun. 2017.

\end{thebibliography}
